# Asymptotic Scaling in the Two-Dimensional $O(3)$ $\sigma$-Model at Correlation Length $10^5$

Sergio Caracciolo
Dipartimento di Fisica
Università di Lecce and INFN – Sezione di Lecce
I-73100 Lecce, ITALIA
Internet: CARACCIO@LE.INFN.IT

Robert G. Edwards
Supercomputer Computations Research Institute
Florida State University
Tallahassee, FL 32306 USA
Internet: EDWARDS@SCRI.FSU.EDU

Andrea Pelissetto
Dipartimento di Fisica and INFN – Sezione di Pisa
Università degli Studi di Pisa
I-56100 Pisa, ITALIA
Internet: PELISSET@SUNTHPI1.DIFI.UNIPI.IT

Alan D. Sokal
Department of Physics
New York University
4 Washington Place
New York, NY 10003 USA
Internet: SOKAL@ACF4.NYU.EDU

November 5, 1994

**Abstract**

We carry out a high-precision Monte Carlo simulation of the two-dimensional $O(3)$-invariant $\sigma$-model at correlation lengths $\xi$ up to $\sim 10^5$. Our work employs a new and powerful method for extrapolating finite-volume Monte Carlo data to infinite volume, based on finite-size-scaling theory. We discuss carefully the systematic and statistical errors in this extrapolation. We then compare the extrapolated data to the renormalization-group predictions. The deviation from asymptotic scaling, which is $\approx 25\%$ at $\xi \sim 10^2$, decreases to $\approx 4\%$ at $\xi \sim 10^5$.

**PACS number(s):** 11.10.Gh, 11.15.Ha, 12.38.Gc, 05.70.Jk

Two-dimensional nonlinear $\sigma$-models are important "toy models" in elementary-particle physics because they share with four-dimensional nonabelian gauge theories the property of perturbative asymptotic freedom [1]. However, the nonperturbative validity of asymptotic freedom has been questioned [2]; and numerical tests of asymptotic scaling in the $O(3)$ $\sigma$-model at correlation lengths $\xi \sim 100$ have shown discrepancies of order 25% [3,4]. In this Letter we employ a new finite-size-scaling extrapolation method [5] (see also Lüscher *et al.* [6] and Kim [7] for related work [8]) to obtain high-precision estimates (errors $\lesssim 2\%$) in the $O(3)$ $\sigma$-model at correlation lengths $\xi$ up to $\sim 10^5$. We find that the discrepancy has decreased to $\approx 4\%$, in good agreement with the asymptotic-freedom predictions.

We study the lattice $\sigma$-model taking values in the unit sphere $S^{N-1} \subset \mathbb{R}^N$, with nearest-neighbor action $\mathcal{H}(\boldsymbol{\sigma}) = -\beta \sum \boldsymbol{\sigma}_x \cdot \boldsymbol{\sigma}_y$. Perturbative renormalization-group computations predict that the (infinite-volume) correlation lengths $\xi^{(exp)}$ and $\xi^{(2)}$ [9] behave as

$$\xi^{\#}(\beta) = C_{\xi^{\#}} e^{2\pi\beta/(N-2)} \left(\frac{2\pi\beta}{N-2}\right)^{-1/(N-2)} \left[1 + \frac{a_1}{\beta} + \frac{a_2}{\beta^2} + \cdots\right] \tag{1}$$

as $\beta \to \infty$. Three-loop perturbation theory yields [11,12]

$$a_1 = -0.014127 + \left(\frac{1}{4} - \frac{5\pi}{48}\right)/(N-2). \tag{2}$$

The nonperturbative constant $C_{\xi^{(exp)}}$ has been computed recently using the thermodynamic Bethe Ansatz [13]:

$$C_{\xi^{(exp)}} = 2^{-5/2} \left(\frac{e^{1-\pi/2}}{8}\right)^{1/(N-2)} \Gamma\left(1 + \frac{1}{N-2}\right). \tag{3}$$

The remaining nonperturbative constant is known analytically only at large $N$ [14]:

$$C_{\xi^{(2)}}/C_{\xi^{(exp)}} = 1 - \frac{0.003225}{N} + O(1/N^2). \tag{4}$$

Previous Monte Carlo studies up to $\xi \sim 100$ agree with these predictions to within about 20–25% for $N = 3$, 6% for $N = 4$ and 2% for $N = 8$ [4,15].

Our extrapolation method [5] is based on the finite-size-scaling Ansatz

$$\frac{\mathcal{O}(\beta, sL)}{\mathcal{O}(\beta, L)} = F_{\mathcal{O}}\big(\xi(\beta, L)/L\,;\,s\big) + O\big(\xi^{-\omega}, L^{-\omega}\big), \tag{5}$$

where $\mathcal{O}$ is any long-distance observable, $s$ is a fixed scale factor (usually $s = 2$), $L$ is the linear lattice size, $F_{\mathcal{O}}$ is a universal function, and $\omega$ is a correction-to-scaling exponent. We make Monte Carlo runs at numerous pairs $(\beta, L)$ and $(\beta, sL)$; we then plot $\mathcal{O}(\beta, sL)/\mathcal{O}(\beta, L)$ versus $\xi(\beta, L)/L$, using those points satisfying both $\xi(\beta, L) \geq$ some value $\xi_{min}$ and $L \geq$ some value $L_{min}$. If all these points fall with good accuracy on a single curve, we choose a smooth fitting function $F_{\mathcal{O}}$. Then, using the functions $F_\xi$ and $F_{\mathcal{O}}$, we extrapolate the pair $(\xi, \mathcal{O})$ successively from $L \to sL \to s^2L \to \ldots \to \infty$. See [5] for how to calculate statistical error bars on the extrapolated values.

We have chosen to use functions $F_{\mathcal{O}}$ of the form

$$F_{\mathcal{O}}(x) = 1 + a_1 e^{-1/x} + a_2 e^{-2/x} + \ldots + a_n e^{-n/x} . \qquad (6)$$

This form is partially motivated by theory, which tells us that $F(x) \to 1$ exponentially fast as $x \to 0$ [16]. Typically a fit of order $3 \leq n \leq 12$ is sufficient; we increase $n$ until the $\chi^2$ of the fit becomes essentially constant. The resulting $\chi^2$ value provides a check on the systematic errors arising from corrections to scaling and/or from inadequacies of the form (6). The discrepancies between the extrapolated values from different $L$ at the same $\beta$ can also be subjected to a $\chi^2$ test. Further details on the method can be found in [5].

We simulated the two-dimensional $O(3)$ $\sigma$-model, using the Wolff embedding algorithm with standard Swendsen-Wang updates [17,18,10]; critical slowing-down appears to be completely eliminated. We ran on lattices $L = 32, 48, 64, 96, 128, 192, 256, 384, 512$ at 180 different pairs $(\beta, L)$ in the range $1.65 \leq \beta \leq 3.00$ (corresponding to $20 \lesssim \xi_\infty \lesssim 10^5$). Each run was between $10^5$ and $5 \times 10^6$ iterations, and the total CPU time was 7 years on an IBM RS-6000/370. The raw data will appear in [19].

Our data cover the range $0.15 \lesssim \xi(L)/L \lesssim 1.0$, and we found tentatively that a tenth-order fit (6) is indicated: see Table 1. Next we took $\xi_{min} = 20$ and sought to choose $L_{min}$ to avoid any detectable systematic error from corrections to scaling. There appear to be weak corrections to scaling ($\lesssim 1.5\%$) in the region $0.3 \lesssim \xi(L)/L \lesssim 0.7$ for lattices with $L \lesssim 64$–96: see the deviations plotted in Figure 1. We therefore investigated systematically the $\chi^2$ of the fits, allowing a different $L_{min}$ for $\xi(L)/L \leq 0.7$ and $> 0.7$: see Table 1. A reasonable $\chi^2$ is obtained when $n \geq 9$ and $L_{min} \geq (128, 64)$. Our preferred fit is $n = 10$ and $L_{min} = (128, 64)$: see Figure 2, where we compare also with the perturbative prediction

$$F_\xi(x; s) = s \left[ 1 - \frac{aw_0 \log s}{2} x^{-2} - a^2 \left( w_1 \log s + \frac{w_0^2 \log^2 s}{8} \right) x^{-4} + O(x^{-6}) \right] \qquad (7)$$

valid for $x \gg 1$, where $a = 1/(N-1)$, $w_0 = (N-2)/2\pi$ and $w_1 = (N-2)/(2\pi)^2$.

The extrapolated values $\xi_\infty^{(2)}$ from different lattice sizes at the same $\beta$ are consistent within statistical errors: only one of the 24 $\beta$ values has a $\chi^2$ too large at the 5% level; and summing all $\beta$ values we have $\chi^2 = 86.56$ (106 DF, level = 92%).

In Table 2 we show the extrapolated values $\xi_\infty^{(2)}$ from our preferred fit and some alternative fits. The discrepancies between these values (if larger than the statistical errors) can serve as a rough estimate of the remaining systematic errors due to corrections to scaling. The statistical errors in our preferred fit are of order 0.2% (resp. 0.7%, 1.1%, 1.6%) at $\xi_\infty \approx 10^2$ (resp. $10^3$, $10^4$, $10^5$), and the systematic errors are of the same order or smaller. The statistical errors at different $\beta$ are strongly positively correlated.

In Figure 3 (points + and ×) we plot $\xi_{\infty,estimate\ (128,64)}^{(2)}$ divided by the two-loop and three-loop predictions (1)–(4). The discrepancy from three-loop asymptotic scaling, which is $\approx 16\%$ at $\beta = 2.0$ ($\xi \approx 200$), decreases to $\approx 4\%$ at $\beta = 3.0$ ($\xi \approx 10^5$).

This is roughly consistent with the expected $1/\beta^2$ corrections. The slight bump at $2.3 \lesssim \beta \lesssim 2.6$ is probably spurious, arising from correlated statistical or systematic errors.

We can also try an "improved expansion parameter" [20,4,12,19] based on the energy $E = \langle \boldsymbol{\sigma}_0 \cdot \boldsymbol{\sigma}_1 \rangle$. First we invert the perturbative expansion [21,12]

$$E(\beta) \;=\; 1 \;-\; \frac{N-1}{4\beta} \;-\; \frac{N-1}{32\beta^2} \;-\; \frac{0.005993(N-1)^2 + 0.007270(N-1)}{\beta^3} \;+\; O(1/\beta^4) \tag{8}$$

and substitute into (1); this gives a prediction for $\xi$ as a function of $1 - E$. For $E$ we use the value measured on the largest lattice; the statistical errors and finite-size corrections on $E$ are less than $5 \times 10^{-5}$, and therefore induce a negligible error (less than 0.5%) on the predicted $\xi$. The corresponding observed/predicted ratios are also shown in Figure 3 (points $\square$ and $\diamond$). The "improved" 3-loop prediction is in excellent agreement with the data.

Let us summarize the conceptual basis of our analysis. The main assumption is that if the Ansatz (5) with a given function $F_\xi$ is well satisfied by our data for $L_{min} \leq L \leq 256$ and $1.65 \leq \beta \leq 3$, then it will continue to be well satisfied for $L > 256$ and for $\beta > 3$. Obviously this assumption could fail, e.g. if [2] at some large correlation length ($\gtrsim 10^3$) the model crosses over to a new universality class associated with a finite-$\beta$ critical point. In this respect our work is subject to the same caveats as any other Monte Carlo work on a finite lattice. However, it should be emphasized that our approach does *not* assume asymptotic scaling [eq. (1)], as $\beta$ plays no role in our extrapolation method. Thus, we can make an unbiased *test* of asymptotic scaling. The fact that we confirm (1) *with the correct nonperturbative constant* (3)/(4) is, we believe, good evidence in favor of the asymptotic-freedom picture. We are unable to imagine how, if there were in fact a finite-$\beta$ critical point [2], the "preasymptotic" region at $\beta \leq 3$ would mimic not only asymptotic freedom but also the nonperturbative constant predicted by the thermodynamic Bethe Ansatz.

Details of this work, including an analysis of the susceptibility $\chi$, will appear elsewhere [19].

We wish to thank Jae-Kwon Kim for sharing his data with us, and for challenging us to push to ever larger values of $\xi/L$. We also thank Steffen Meyer, Adrian Patrascioiu, Erhard Seiler and Ulli Wolff for helpful discussions. The authors' research was supported by CNR, INFN, DOE contracts DE-FG05-85ER250000 and DE-FG05-92ER40742, NSF grant DMS-9200719, and NATO CRG 910251.

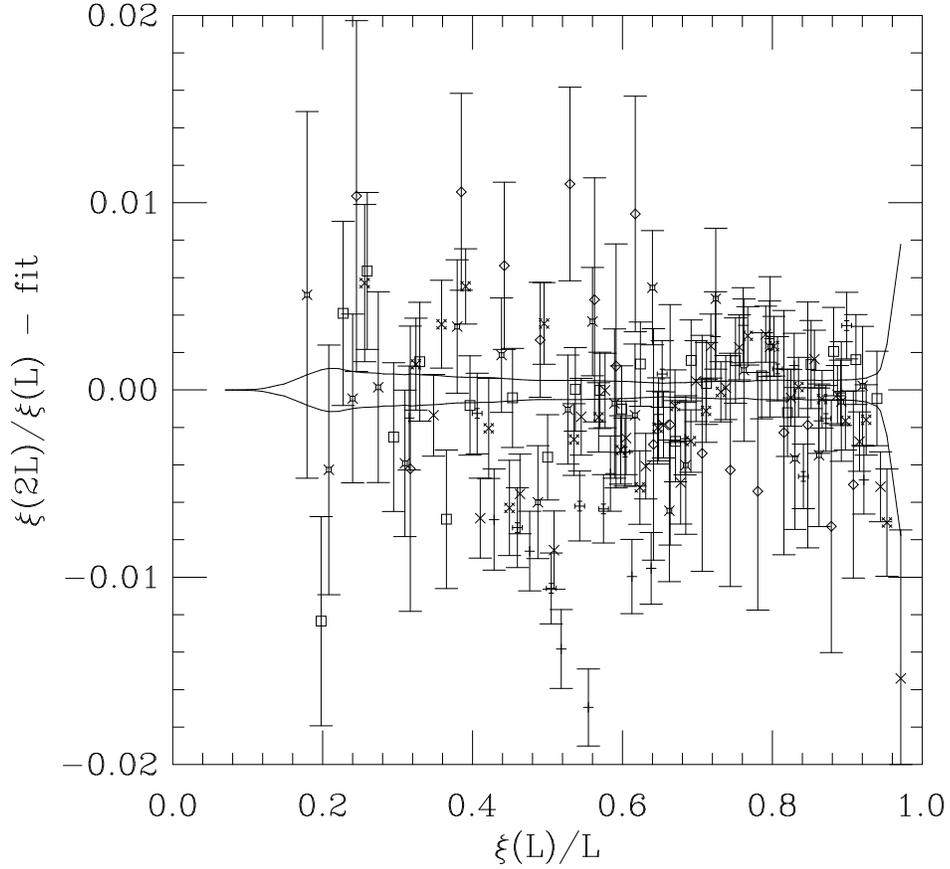

Figure 1: Deviation of points from fit to $F_\xi$ with $s = 2$, $\xi_{min} = 20$, $L_{min} = 128$, $n = 10$. Symbols indicate $L = 32$ (+), 48 (⊞), 64 (×), 96 (⋇), 128 (□), 192 (⋈), 256 (◇). Error bars are one standard deviation. Curves near zero indicate statistical error bars (± one standard deviation) on the function $F_\xi(x)$.

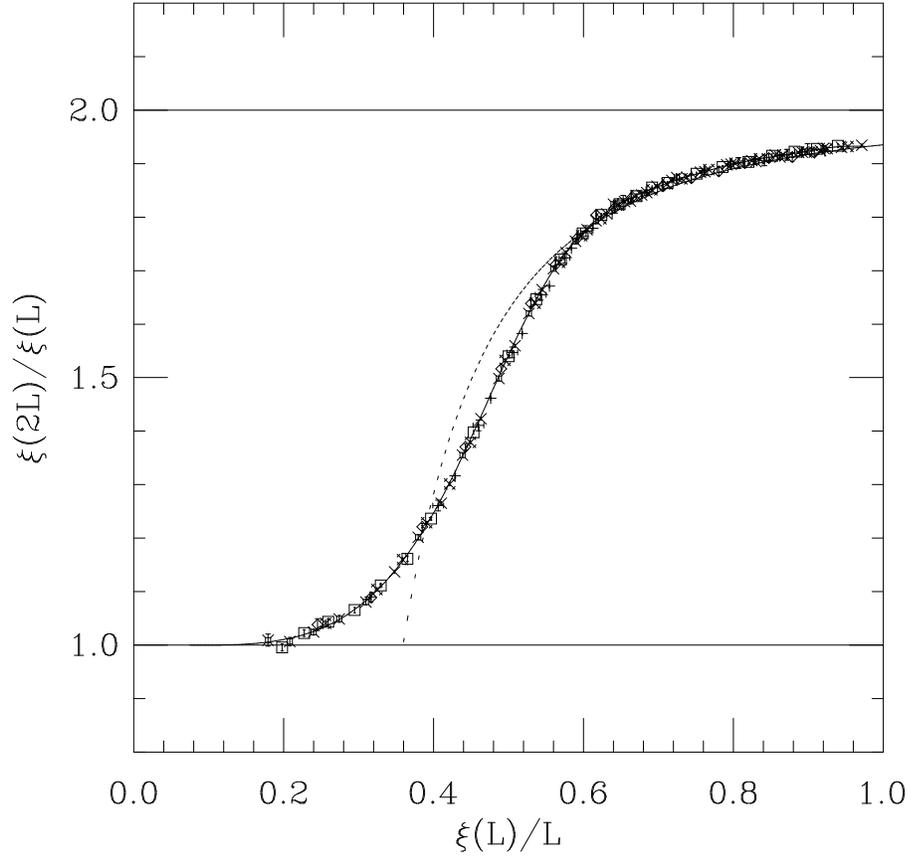

Figure 2: $\xi(\beta, 2L)/\xi(\beta, L)$ versus $\xi(\beta, L)/L$. Symbols indicate $L = 32$ (+), 48 (⊞), 64 (×), 96 (✳), 128 (□), 192 (⋈), 256 (◇). Error bars are one standard deviation. Solid curve is a tenth-order fit in (6), with $\xi_{min} = 20$ and $L_{min} = 128$ (resp. 64) for $\xi(L)/L \leq 0.7$ (resp. $> 0.7$). Dashed curve is the perturbative prediction (7).

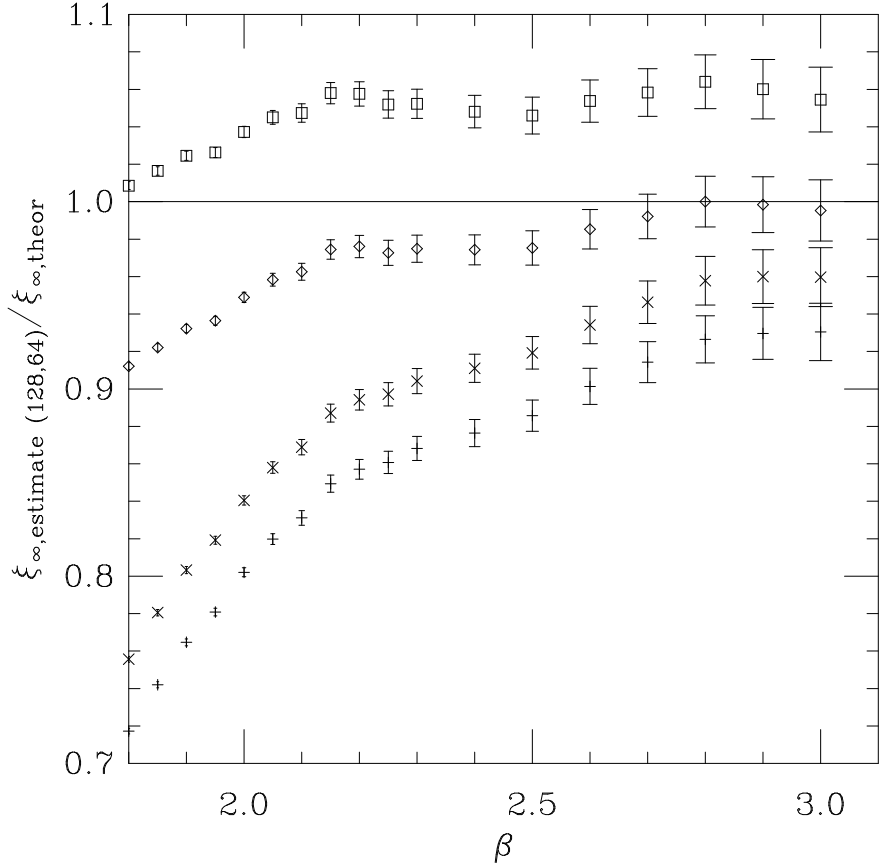

Figure 3: $\xi^{(2)}_{\infty,estimate\ (128,64)}/\xi^{(2)}_{\infty,theor}$ versus $\beta$. Error bars are one standard deviation (statistical error only). There are four versions of $\xi^{(2)}_{\infty,theor}$: standard perturbation theory in $1/\beta$ gives points $+$ (2-loop) and $\times$ (3-loop); "improved" perturbation theory in $1 - E$ gives points $\square$ (2-loop) and $\diamond$ (3-loop).

| $L_{min}$ | DF | $n=7$ | $n=8$ | $n=9$ | $n=10$ | $n=11$ | $n=12$ |
|---|---|---|---|---|---|---|---|
| (64,64) | $108-n$ | 278.38 0.0% | 183.80 0.0% | 144.34 0.2% | 137.82 0.5% | 135.77 0.6% | 135.01 0.5% |
| (96,32) | $107-n$ | 228.85 0.0% | 164.46 0.0% | 129.38 1.9% | 124.87 3.0% | 122.15 3.7% | 120.48 4.0% |
| (96,64) | $97-n$ | 207.32 0.0% | 137.18 0.1% | 108.23 7.1% | 103.13 11.4% | 102.02 11.5% | 101.59 10.6% |
| (96,96) | $87-n$ | 190.61 0.0% | 115.05 0.5% | 100.99 4.1% | 93.90 9.2% | 93.89 8.0% | 93.73 7.1% |
| (128,32) | $93-n$ | 160.17 0.0% | 121.29 0.6% | 99.35 12.1% | 94.82 17.7% | 94.20 16.8% | 86.65 31.3% |
| (128,64) | $83-n$ | 139.60 0.0% | 95.94 5.2% | 78.23 34.6% | 72.91 48.1% | 72.89 44.9% | 68.43 56.4% |
| (128,96) | $73-n$ | 126.20 0.0% | 79.03 11.3% | 71.12 25.3% | 64.33 43.0% | 63.29 43.1% | 59.72 52.2% |
| (128,128) | $64-n$ | 101.05 0.0% | 63.45 23.1% | 61.96 24.2% | 59.70 27.6% | 59.28 25.7% | 52.89 43.9% |
| (192,32) | $75-n$ | 110.42 0.1% | 93.41 1.8% | 76.13 18.5% | 70.61 29.6% | 65.15 43.6% | 62.16 50.6% |
| (192,64) | $65-n$ | 90.60 0.4% | 69.57 12.3% | 55.03 51.1% | 47.60 75.0% | 45.12 80.0% | 43.74 81.4% |
| (192,96) | $57-n$ | 82.54 0.3% | 55.94 23.0% | 49.49 41.4% | 38.90 79.4% | 38.67 77.0% | 37.53 77.8% |

Table 1: $\chi^2$ and confidence level for the fit (6) of $\xi(\beta,2L)/\xi(\beta,L)$ versus $\xi(\beta,L)/L$. DF = number of degrees of freedom. The first (resp. second) $L_{min}$ value applies for $\xi(L)/L \leq 0.7$ (resp. $> 0.7$). In all cases $\xi_{min} = 20$.

| $L_{min}$ | 1.90 | 1.95 | 2.00 | 2.05 | 2.10 | 2.15 | 2.20 | 2.25 |
|---|---|---|---|---|---|---|---|---|
| (96,64) | 122.43 (0.25) | 166.79 (0.36) | 228.37 (0.55) | 311.54 (0.93) | 420.52 (1.59) | 574.16 (2.51) | 774.24 (3.69) | 1039.1 ( 5.7) |
| (96,96) | 122.55 (0.25) | 166.95 (0.37) | 228.93 (0.57) | 312.29 (0.93) | 421.61 (1.63) | 574.96 (2.51) | 776.03 (3.73) | 1038.2 ( 5.5) |
| (128,32) | 122.34 (0.29) | 166.68 (0.42) | 228.50 (0.66) | 311.84 (1.09) | 422.67 (1.94) | 577.41 (3.13) | 779.33 (4.80) | 1048.9 ( 7.3) |
| *(128,64)* | *122.34 (0.29)* | *166.66 (0.43)* | *228.54 (0.67)* | *311.99 (1.10)* | *422.73 (1.97)* | *577.73 (3.12)* | *780.04 (4.76)* | *1048.7 ( 7.3)* |
| (128,96) | 122.25 (0.29) | 166.54 (0.43) | 228.11 (0.66) | 311.59 (1.10) | 421.71 (1.90) | 576.52 (3.06) | 778.40 (4.58) | 1045.9 ( 7.3) |
| (128,128) | 122.36 (0.29) | 166.68 (0.43) | 228.59 (0.69) | 312.06 (1.13) | 422.89 (2.00) | 577.94 (3.09) | 781.23 (4.79) | 1046.7 ( 7.3) |
| (192,32) | 122.40 (0.40) | 166.95 (0.60) | 229.05 (0.93) | 312.94 (1.49) | 424.90 (2.69) | 580.40 (4.39) | 784.04 (7.14) | 1057.7 (11.4) |
| (192,64) | 122.41 (0.38) | 166.94 (0.57) | 229.15 (0.90) | 312.86 (1.44) | 425.42 (2.62) | 580.91 (4.41) | 785.39 (7.11) | 1057.3 (11.2) |
| (192,96) | 122.43 (0.39) | 167.02 (0.58) | 229.30 (0.90) | 313.23 (1.45) | 426.08 (2.70) | 581.91 (4.44) | 787.63 (7.18) | 1057.9 (11.2) |
| Kim | 122.0 ( 2.7) | — | 227.8 ( 3.2) | 306.6 ( 3.9) | 419 ( 5) | 574 ( 8) | 766 ( 7) | — |

| $L_{min}$ | 2.30 | 2.40 | 2.50 | 2.60 | 2.70 | 2.80 | 2.90 | 3.00 |
|---|---|---|---|---|---|---|---|---|
| (96,64) | 1403.4 ( 8.3) | 2539.1 (17.9) | 4619.7 (38.6) | 8460.1 ( 81.7) | 15499 (172) | 28413 (362) | 51624 ( 746) | 93601 (1475) |
| (96,96) | 1402.0 ( 8.4) | 2541.5 (19.2) | 4605.9 (44.5) | 8450.7 (101.2) | 15401 (218) | 28119 (455) | 51356 ( 934) | 93641 (1923) |
| (128,32) | 1416.7 (10.6) | 2566.2 (20.8) | 4687.7 (41.3) | 8559.0 ( 81.1) | 15594 (161) | 28622 (322) | 51955 ( 651) | 94133 (1345) |
| *(128,64)* | *1416.8 (10.5)* | *2568.8 (21.2)* | *4671.7 (43.9)* | *8569.0 ( 91.6)* | *15690 (189)* | *28737 (389)* | *52189 ( 779)* | *94643 (1554)* |
| (128,96) | 1414.1 (10.8) | 2558.1 (22.8) | 4628.6 (48.4) | 8478.0 (104.3) | 15507 (226) | 28360 (470) | 51695 ( 961) | 94033 (1930) |
| (128,128) | 1415.5 (11.2) | 2572.1 (26.2) | 4637.7 (62.0) | 8437.2 (143.2) | 15336 (311) | 27947 (666) | 51319 (1392) | 94627 (2922) |
| (192,32) | 1425.3 (17.0) | 2584.4 (32.8) | 4716.9 (62.6) | 8622.7 (118.4) | 15638 (225) | 28820 (432) | 52345 ( 843) | 94724 (1660) |
| (192,64) | 1427.6 (17.0) | 2582.2 (32.9) | 4702.7 (62.7) | 8625.0 (123.1) | 15802 (244) | 28952 (482) | 52502 ( 934) | 95266 (1819) |
| (192,96) | 1427.0 (17.0) | 2584.2 (33.4) | 4688.2 (65.4) | 8599.8 (133.7) | 15660 (269) | 28663 (542) | 52314 (1082) | 95304 (2163) |
| Kim | 1402 ( 22) | 2499 ( 41) | 4696 ( 128) | 8022 ( 234) | 15209 (449) | — | — | — |

Table 2: Estimated correlation lengths $\xi_\infty^{(2)}$ as a function of $\beta$, from various extrapolations. Error bar is one standard deviation (statistical errors only). All extrapolations use $s = 2$, $\xi_{min} = 20$ and $n = 10$. The first (resp. second) $L_{min}$ value applies for $\xi(L)/L \leq 0.7$ (resp. $> 0.7$). Our preferred fit is $L_{min} = (128, 64)$, shown in italics. Kim is the estimate from [7].